# Crystal Structure Determination via Inverse EXAFS Analysis: A Comparative Study Utilizing the Demeter Software Package

O. Murat Ozkendir[1*]

1. *Dept. of Natural and Mathematical Sciences, Faculty of Engineering, Tarsus University, 33400,Tarsus, Mersin, Turkiye*

**Abstract**

This study introduces a novel approach for crystal structure analysis, utilizing Inverse EXAFS Analysis (IEA). To assess the reliability of IEA, we applied it to various experimentally studied materials, including $LiCrO_2$ and $CuFeO_2$. Our findings demonstrate that IEA offers a promising alternative to traditional techniques like XRD, particularly in cases where instrumentation or crystal structure defects pose challenges. IEA effectively revealed the crystal structures of both $LiCrO_2$ and $CuFeO_2$, demonstrating its ability to accurately characterize complex materials. The technique's potential to enhance XAFS data analysis is significant, providing researchers with a valuable tool for crystal structure determination. Future developments in IEA could further expand its capabilities and make it a more accessible and efficient method for materials scientists.

*Corresponding author: ozkendir@gmail.com

1. Introduction

With the growing demand for new, high-performance, and skillful novel materials that exhibit highly desirable features in achieving ultimate technological consequences, assign procedures that result in more precise conclusions. Crystal structure plays an important function in materials investigation and determination [1-3]. The feature of crystal structure where x-ray diffraction (XRD) is the most popular, among others, has a vital role [4]. However, the technique's success puts other features to the forefront, such as the quality of the instrumentation or measuring system, the data gathering method, or the crystal structure's solidity. Sometimes single crystal material data cannot be processed due to instrumentation failure, or instruments may fail due to crystal structure defects in sample preparation methods that are not properly planned. In such cases, scientists begin looking for a different approach to detect the crystal structure of the samples under research in order to precisely specify its properties. The crystal structure of a sample determines its properties, such as electrical, electronic, and magnetic, as well as the purpose for which it was prepared and the expected results. In such a circumstance, various methodologies may provide useful data in forming an opinion regarding the crystal structure content based on its atomic types. However, it is pleasing to report that there is now a precise choice for determining crystal structure attributes without employing crystal structure data giving techniques such as XRD or Van Laue, or vice versa, but rather EXAFS, the Inverse EXAFS Analysis (IEA).

EXAFS, or Extended X-ray Absorption Fine Structure, is derived from the oscillatory features observed in the high-energy region (extended region) of an X-ray absorption spectrum (**Figure 1**). These oscillations arise from the interference between the incoming and outgoing photoelectron wavefunctions emitted from the absorbing atom when excited by an X-ray photon.

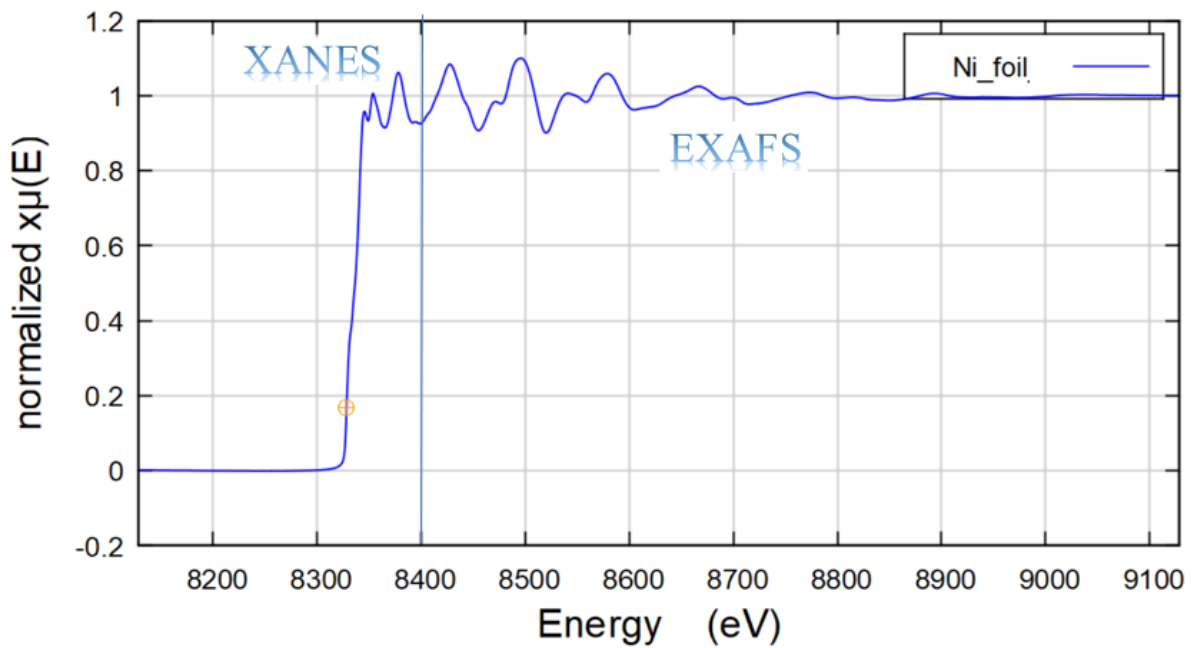

**Figure 1.** X-ray Absorption Fine Structure (XAFS) Spectrum with Indicated Regions

When an X-ray photon excites an inner-shell electron to a higher energy state above the Fermi level, an absorption event occurs. This process is known as inner-shell excitation. If the photon's energy exceeds the electron's binding energy, the excess energy is transferred to the electron, which is then ejected from the atom as a photoelectron. The photoelectron travels through the surrounding atomic medium. As a photoelectron approaches an atom within a crystal, it experiences both attractive forces from the nucleus and repulsive forces from the outer-shell electrons due to Coulombic interactions. These interactions cause the photoelectron's path to deviate, a phenomenon known as 'scattering.' Scattering can also occur due to the potential energy landscape created by the arrangement of atoms in the crystal lattice. When the photoelectron's kinetic energy is depleted, it is recaptured by the atom from which it originated, effectively 'dying' and contributing to the electron population of the crystal.

The observed photoelectrons' tail component fluctuations, which are positive when the photoelectrons' wave functions interact in phase and negative when out of phase with x-rays, can be recovered using the equation below:

$$\chi(E) = \frac{\mu(E) - \mu_0(E)}{\mu_0(E)} \qquad (1)$$

Here, $\mu_0(E)$ is the background absorption (or attenuation) from an isolated absorber, and $\mu(E)$ is the absorbance in the material. The applied formula on the XAFS data provides the scattering intensities like the data given in **Figure 2**.

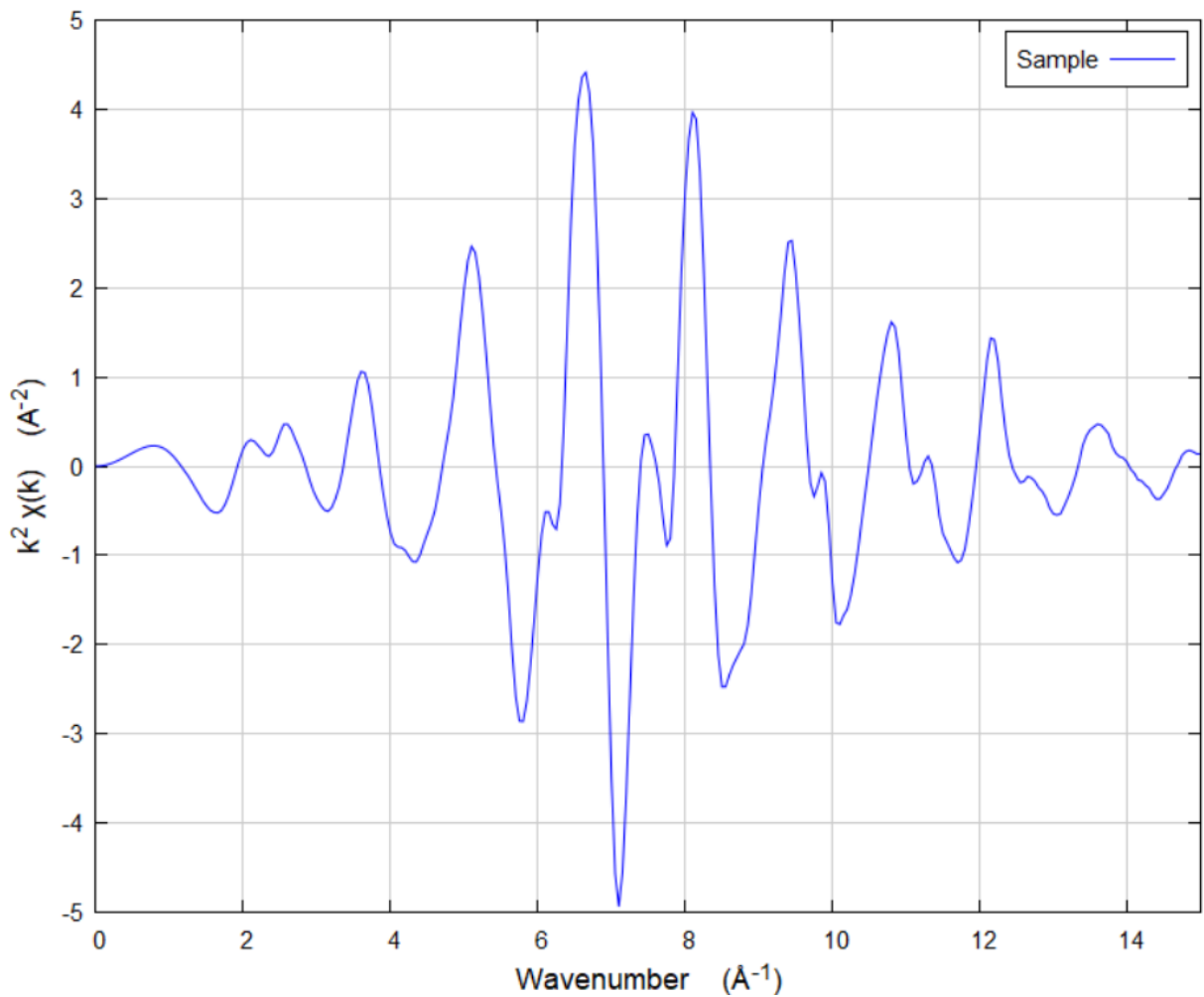

**Figure 2.** EXAFS scattering data of a studied sample material

Fourier transformation of the EXAFS oscillations provides information about the distances between the absorbing atom and neighboring atoms.

This study aims to demonstrate the application of Inverse EXAFS (Fitting) Analysis (IEA) for crystal structure determination, a technique complementary to Rietveld refinement of diffraction patterns [2, 5-6]. The theoretical approach involves using the FEFF code within the Demeter software package for EXAFS fitting, incorporating Athena, Artemis, and Hephaestus modules for data analysis (Figure 3). This analysis provides insights into the local atomic structure of the sample material. [7, 8]. FEFF is an automated program for ab initio multiple scattering calculations of X-ray Absorption Fine Structure (XAFS), X-ray Absorption Near-Edge Structure (XANES) and various other spectra for clusters of atoms. Demeter is a comprehensive system for processing and analyzing X-ray Absorption Spectroscopy data.

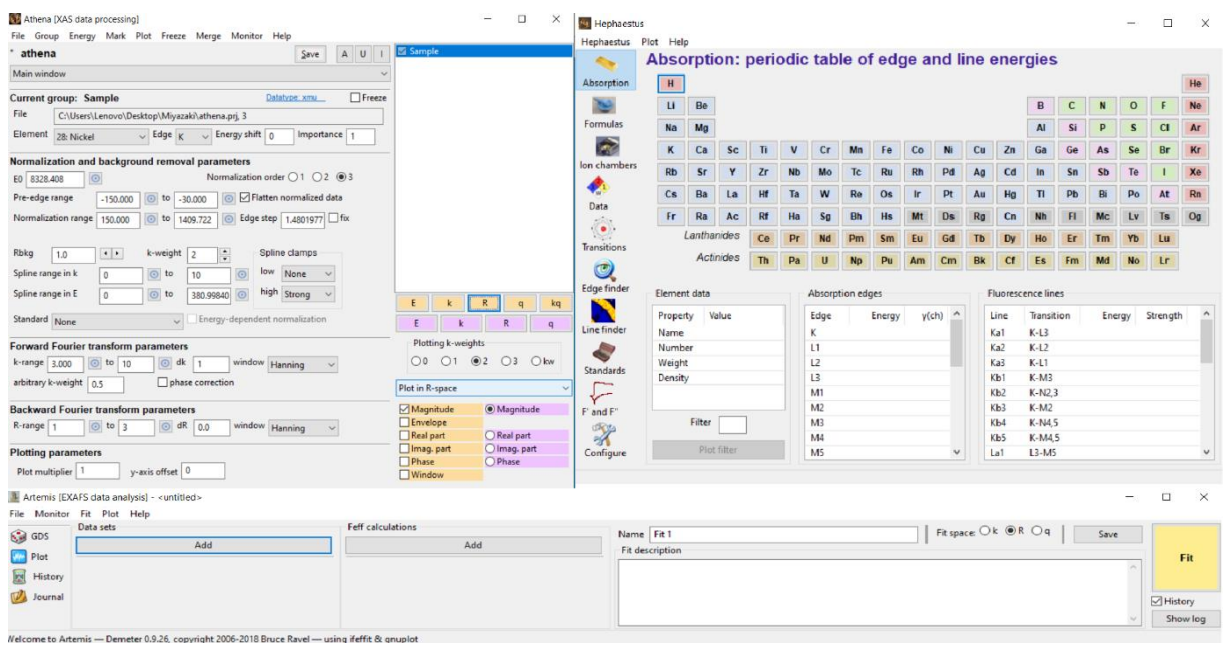

**Figure 3.** Demeter Software Package Modules

Within the Demeter software package, the Athena module is primarily used for general XAFS data processing and EXAFS data extraction. The Artemis module is specifically designed for detailed EXAFS fitting analysis. Additionally, the Hephaestus module provides support for various general analysis tasks.

The motivation for this study arose from the challenges encountered in obtaining crystallographic data via powder X-ray diffraction (PXRD) during the pandemic, which restricted international travel and access to synchrotron facilities. These limitations necessitated the exploration of alternative approaches for crystal structure determination. The arduous process of data analysis, often requiring extensive time and effort, can yield valuable insights, highlighting the potential benefits of perseverance in scientific research. This study compares the results obtained from two different samples, whose XAFS data were collected at separate synchrotron facilities and analyzed independently. The results obtained from IEA were corroborated by XRD data collected at different times and using various instruments. While numerous samples were analyzed with consistent results, two representative cases are presented here for detailed examination. Among the studied materials, $LiCrO_2$ and $CuFeO_2$ (incorrectly identified by researchers) were selected to demonstrate the effectiveness of the IEA technique.

**Materials and Method**

**Sample 1: $LiCrO_2$ Material**

The researchers aimed to synthesize $LiCrO_2$, a layered cathode material with trigonal rhombohedral geometry (**Figure 4**). Despite its relatively weak cathode properties, $LiCrO_2$ has garnered interest due to the redox activity of chromium during electrochemical reactions ($Cr^{3+}$ to $Cr^{6+}$) [9]. Believing they had successfully prepared $LiCrO_2$, the researchers initiated their analysis by extracting XAFS data from the chromium K-edge of the sample, and this study was another good chance to confirm the reliability of the IEA technique.

The initial step in a typical EXAFS analysis using the Demeter package involves extracting the scattering data from the XAFS spectrum using the Athena module. While the

XANES region, which provides valuable insights into chemical bonding and electronic interactions, can also be processed using Athena or other specialized software, the focus of this study is on the EXAFS analysis. The extracted data is then saved for further analysis using the Artemis module. The Hephaestus module is instrumental in preparing the data for accurate calibration and other preprocessing tasks, facilitating in-depth analysis using Athena or other relevant software. **Figure 4** illustrates a typical processing workflow for a sample (not related to this study).

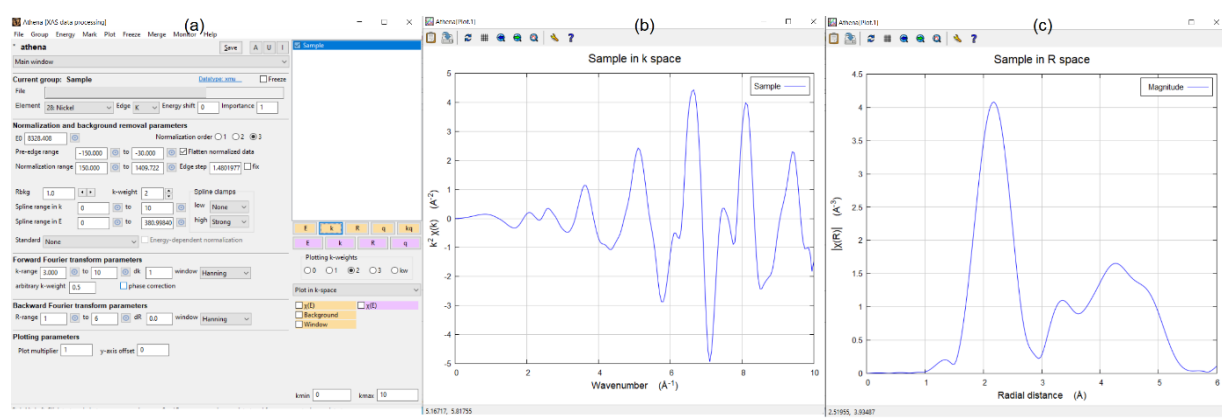

**Figure 4.** a) EXAFS Data Extraction using Athena b) Scattering Data c) Fourier-Transformed Radial Distribution Function

The standard technique of an EXAFS data fitting and diversion for Inverse EXAFS Analysis (IEA) is detailed as follows:

i. **EXAFS Data Extraction and Fitting**

In a typical EXAFS analysis, researchers extract scattering data from an XAFS spectrum using Athena or a similar software. To determine atomic types, coordination numbers, and bond lengths, EXAFS fitting is performed using a crystal structure file (CIF) containing crystallographic information (**Figure 5**). The EXAFS fitting process incorporates the experimentally determined lattice parameters, which can be obtained through methods such as

Rietveld refinement. The determined crystal structure parameters are used to refine the EXAFS fitting process, ensuring accurate determination of atomic types and locations. A high-quality fit between the experimental data and the theoretical calculations generated by the FEFF code within Artemis indicates the correctness of the analysis. As shown in **Figure 6**, the resulting data card provides information on atomic types, coordination numbers, and precise bond lengths.

**Figure 5.** EXAFS Fitting Using Crystallographic Information File (CIF)

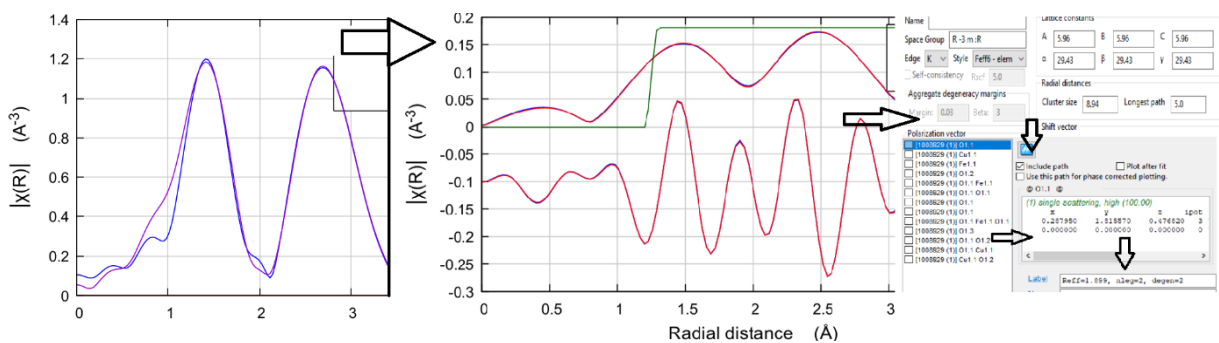

**Figure 6.** EXAFS Fitting Comparison: Experimental Data vs. Theoretical Calculation

In a typical EXAFS analysis, prior knowledge of the crystal structure is essential. However, the question often arises: 'What if crystal structure data is unavailable before performing EXAFS analysis?' or 'Can EXAFS data be used to determine crystal structure parameters?' The answer is affirmative. By employing Inverse EXAFS Analysis (IEA), a technique that involves fitting the EXAFS data in reverse, it is possible to extract valuable information about the crystal structure.

### ii.     The Inverse EXAFS Analysis (IEA)

While crystal structure properties are typically determined using techniques like X-ray diffraction, the initial step in IEA is analogous. The first step involves identifying the most probable crystal structure for the sample by comparing experimental and calculated EXAFS spectra. **Figure 7** illustrates the comparison of calculated EXAFS spectra for various potential crystal structures obtained from databases such as the Crystallography Open Database (COD; https://www.crystallography.net/cod/) or the Materials Project (https://next-gen.materialsproject.org/).

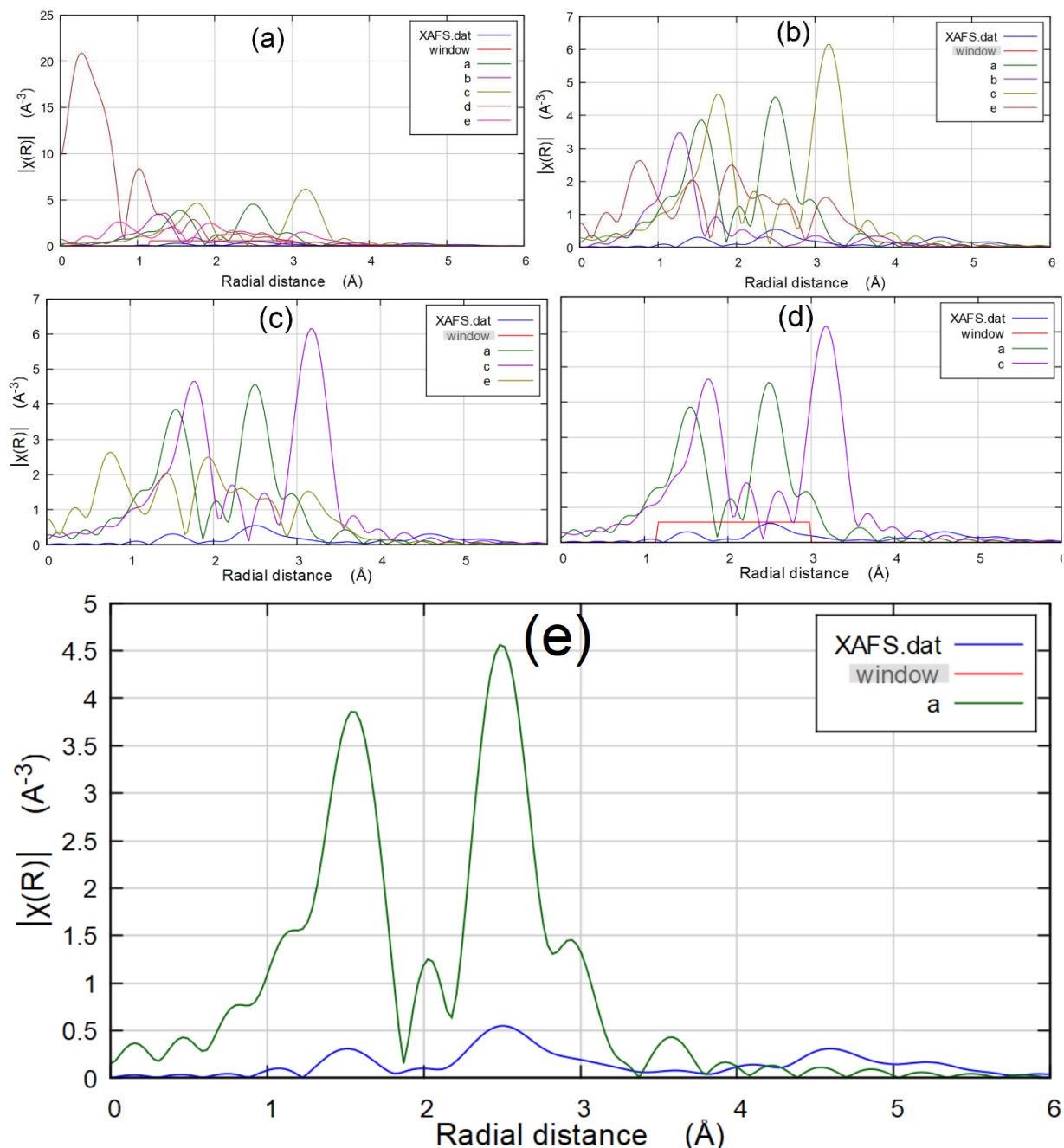

**Figure 7.** Comparison of Crystal Structures and Corresponding EXAFS Fits

A successful IEA study requires downloading CIF files for all potential crystal structures containing the elements present in the sample, not just the desired structure. To identify the best-matching crystal structure(s), it is essential to compare the experimental FT-EXAFS data with calculated FT-EXAFS data for all candidate structures. In the case of $LiCrO_2$, CIF files were obtained from the COD and Materials Project databases, resulting in a list of five distinct crystal structures containing Li, Cr, and O. **Table 1** provides the crystal structure properties and their corresponding alphabetic codes, as shown in **Figure 6**. The 'XAFS.dat' file represents the experimental XAFS data collected at the synchrotron facility and subjected to analysis.

**Table 1.** List of crystal structures used in the comparison

| Code | Crystal Name | Geometry/Space Group | Lattice Parameters | Angles |
|---|---|---|---|---|
| **_a_** | $LiCrO_2$ | *Trigonal/ R -3 m :H* | a=b=2.91Å, c=13.82 Å | α=β=90°, γ=120° |
| **_b_** | $Li_2CrO_4$ | *Trigonal/ R -3:H* | a=b=14.005 Å, c=9.405 Å | α=β=90°, γ=120° |
| **_c_** | $LiCrO_2$ | *Cubic/F m -3 m* | a=b=c=5.00085 Å | α=β=γ=90° |
| **_d_** | $LiCrO_2$ | *Hexagonal/ P 63 m c* | a=b=2.9535 Å, c=9.977 Å | α=β=90°, γ=120° |
| **_e_** | $LiCrO_2$ | *Tetragonal/ I 41/ a m d* | a=b=5.739 Å, c=5.244 Å | α=β=γ=90° |

To identify the most suitable crystal structure, calculated files should be systematically eliminated by comparing their peak structures to the experimental data. Begin by discarding files with significantly different peak patterns. Gradually remove files with increasingly similar peak structures, observing the changes in the fit until a dominant crystal structure emerges. For clarity, each calculated path can be added to the graph with a unique label, as illustrated in **Figure 7**. If a perfect match cannot be achieved after eliminating calculated files, the two most similar structures should be combined and compared to the best-fitting structure to identify the dominant crystal component. This process is crucial to account for potential polycrystallinity in the material. In the case of $LiCrO_2$, five crystal structures containing Li, Cr, and O were eliminated from the COD and Materials Project databases. As shown in **Figure 7.d**, the two remaining structures with the highest similarity were trigonal 'a' and cubic 'c'. Ultimately, the trigonal *'R -3 m'* crystal structure of $LiCrO_2$ was determined to be the most dominant component. There are two primary approaches to proceed with the analysis: 1) manually adjusting lattice parameters to achieve a match between experimental and calculated peak positions before initiating fitting, or 2) directly starting the fitting process and making iterative adjustments. The first approach, while potentially time-consuming, can yield more precise results. By systematically varying lattice parameters and comparing the resulting peak shifts (**Figure 8**), researchers can iteratively refine the crystal structure until a satisfactory convergence is achieved.

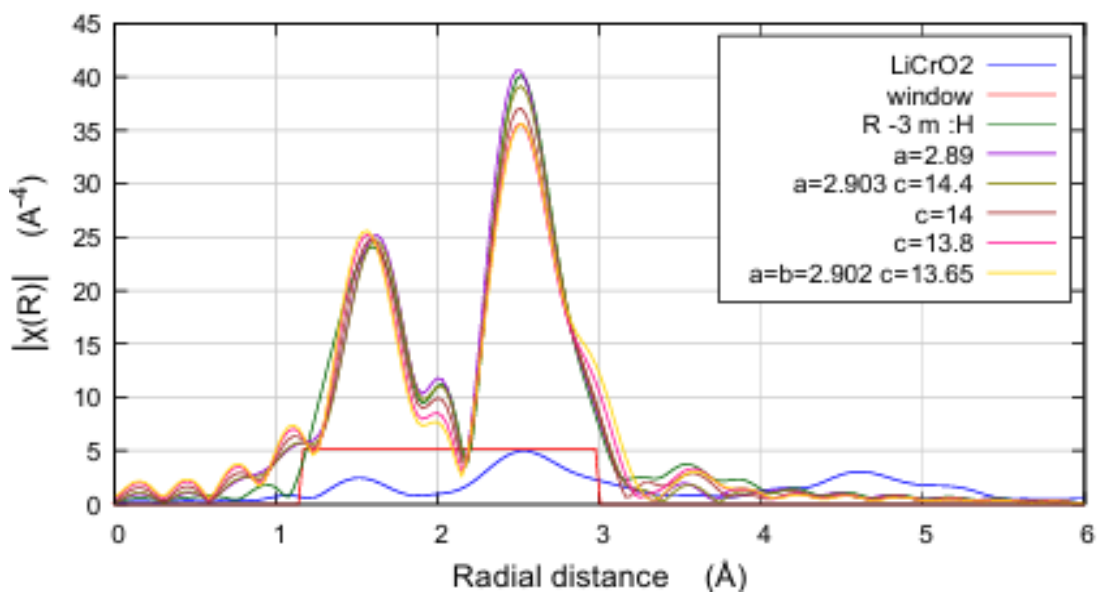

**Figure 8.** Iterative Lattice Parameter Refinement for Peak Position Matching

An alternative approach involves directly initiating the fitting process and extracting the refined crystal lattice parameters from the fitting log file. This iterative process continues until a satisfactory match between experimental and calculated spectra is achieved. A preliminary fitting step is performed, generating a log file that contains crucial information. The 'guess card' within the fitting parameters, highlighted by a red rectangle frame in **Figure 9**, plays a pivotal role in this process. The 'guess card' within the fitting parameters, using carefully selected and accurate values, generates a log file that indicates the convergence status and associated penalties. Key parameters to monitor include the $SO_2$ amplitude, Debye temperature, delta of initial energy, effective R values, and delta R factors. With each iteration, the log file displays convergence values, including the 'R' value and 'delta R' values, which provide insights into the optimal bond lengths within the crystal structure. When a parameter reaches a scaling factor of zero, it suggests that the current value represents the best fit for the given lattice parameters. However, this does not necessarily mean that the exact crystal parameters have been determined. The ideal 'R' value for the atoms (indicated by a green arrow) should be used to refine the lattice parameters in the CIF file through systematic adjustments. A valuable tip is to repeat this process multiple times on the same CIF file without uploading it to the Artemis layout until the desired fitting path is achieved (indicated by a blue arrow and frame).

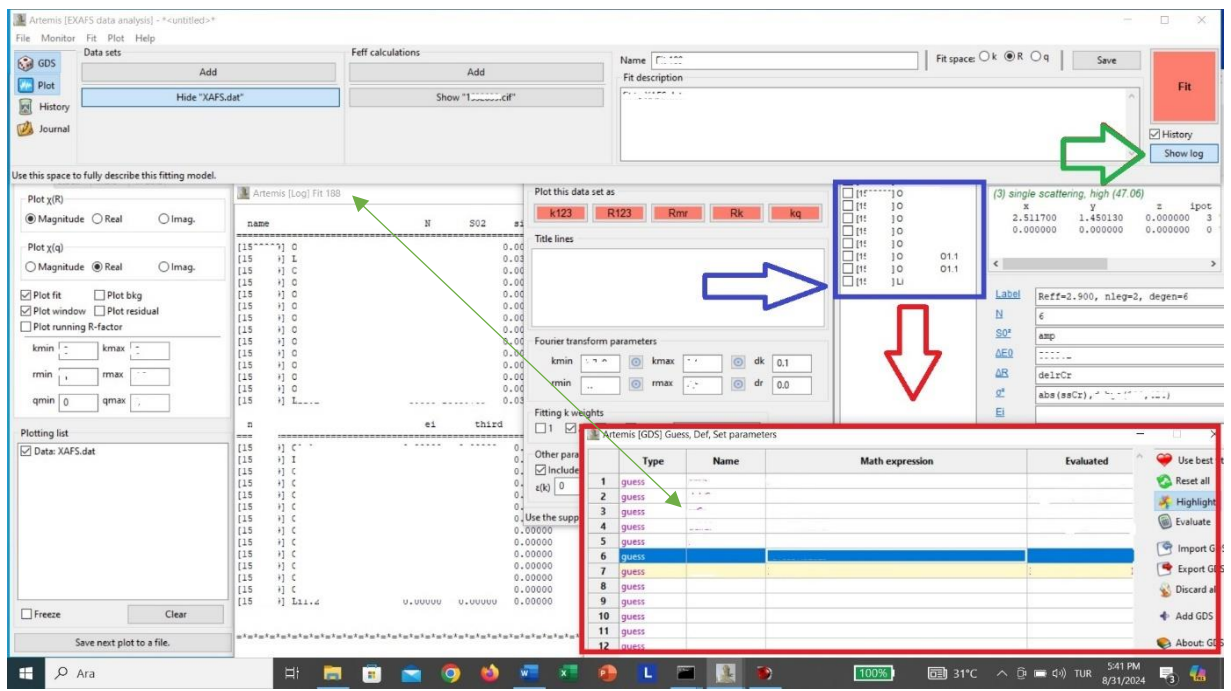

**Figure 9.** Iterative Fitting and Parameter Refinement

When the atomic positions have converged to their optimal values, you will observe no further changes in the guess parameters and/or the error margins on the right side of the panel. The progress of the fitting process, including the applied parameters and the refined atomic positions from the log files, is illustrated in **Figure 10**. The narrower R-value scale indicates the successful convergence of the fitting process.

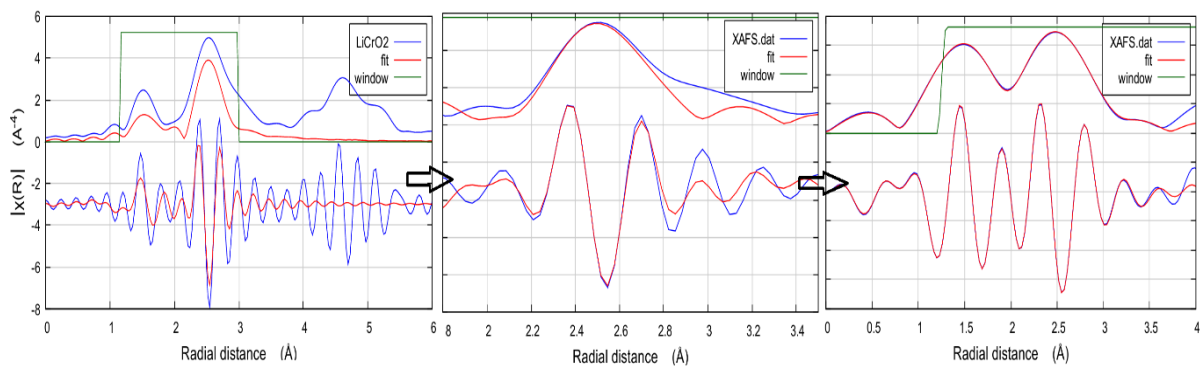

**Figure 10.** Iterative Fitting and Parameter Refinement

By refining the lattice parameters based on the distances determined from the final log file, and iteratively applying the fitting procedure, you can achieve convergence with a zero delta R value, indicating that the optimal lattice parameters for the studied material have been obtained. **Figure 11** demonstrates the results of this process for $LiCrO_2$, where the determined lattice parameters were used to refine the XRD pattern.

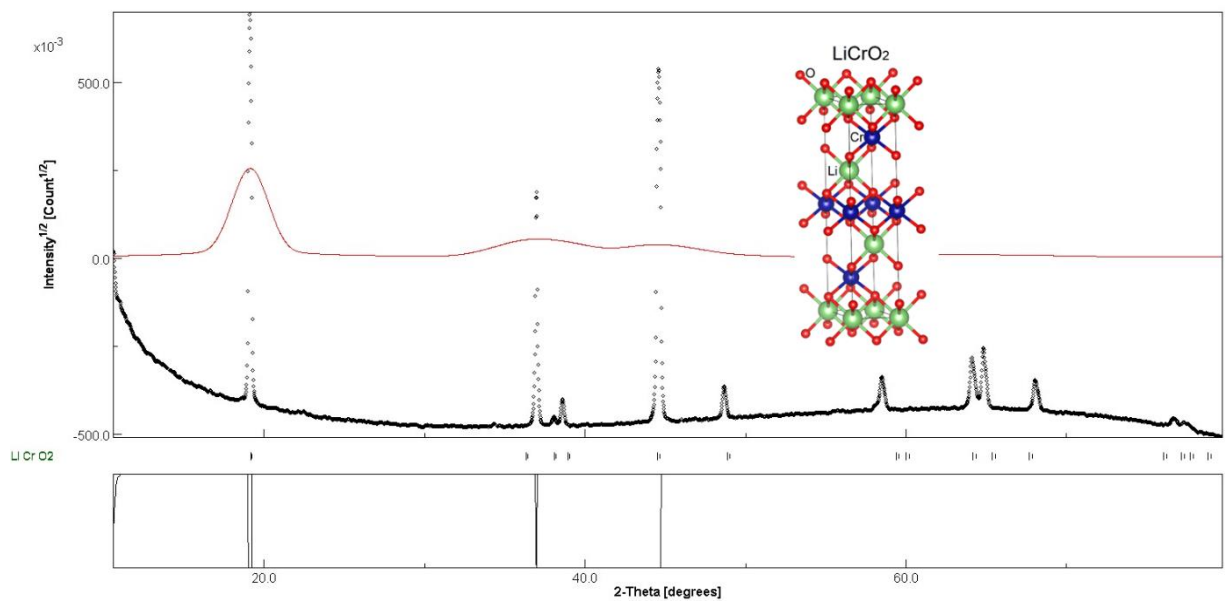

**Figure 11.** Refinement of Lattice Parameters Using Rietveld Analysis

The refined lattice parameters obtained from EXAFS fitting exhibited excellent agreement with the XRD pattern of the studied material for trigonal "*R -3 m: H*" crystal, confirming the researchers XRD analysis.

**Sample 2: $CuFeO_2$ Material**

$CuFeO_2$ (CFO) is a p-type thermoelectric oxide composed entirely of abundant and non-toxic elements, conforming to the chemical formula $ABO_2$. In this structure, monovalent Cu cations occupy the A-site, while the B-site is occupied by a trivalent element such as Cr, Fe, or Al. CFO is a promising magnetic semiconductor material exhibiting antiferromagnetic ordering at a Néel temperature ($T_N$) of 11 K [10]. The major crystal structures of $CuFeO_2$ are depicted in **Figures 12a** and **12b**.

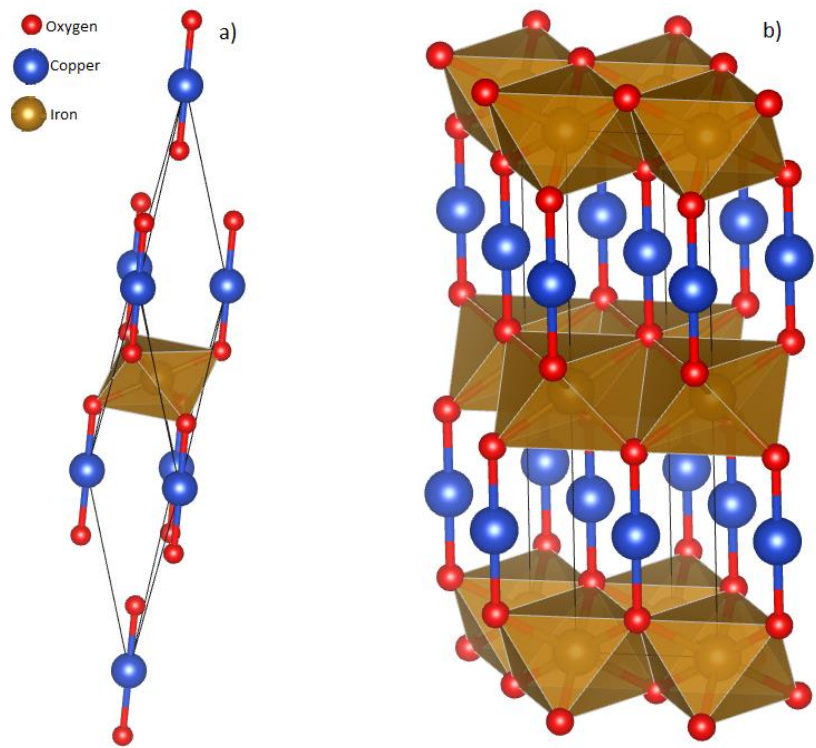

**Figure 12.** a) Trigonal (rhombohedral) $CuFeO_2$ crystal structure b) Hexagonal $CuFeO_2$ crystal structure [11]

XAFS spectra were collected at room temperature from the Fe K-edge of the $CuFeO_2$ sample. The analysis followed the same procedures described previously (**Figure 13**), i.e. elimination of the non resembling data.

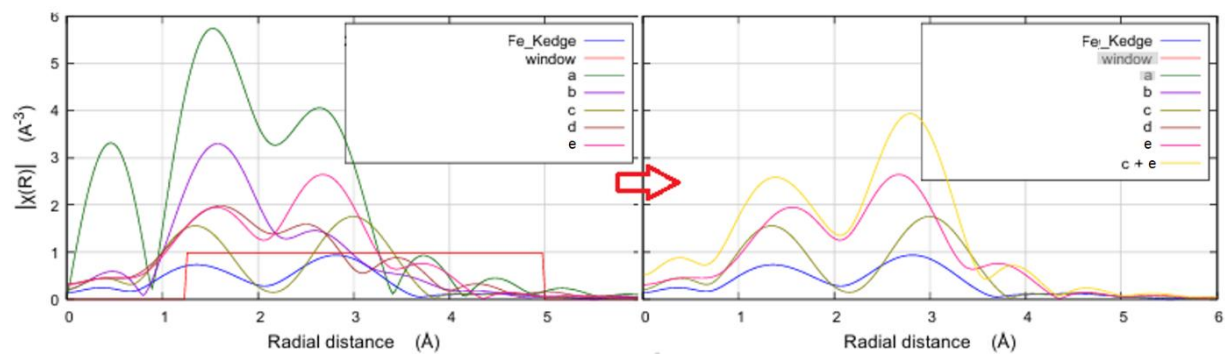

**Figure 13.** Preliminary EXAFS Fitting for $CuFeO_2$ Crystal Structure

To assess crystal structure similarity based on FT-EXAFS data, six distinct crystal structures were obtained from the COD and Materials Project databases, as given in **Table 2**. Crystal structures with identical features were eliminated, leaving six candidates for preliminary EXAFS fitting.

**Table 2.** List of crystal structures used in the comparison for CFO material

| Code | Crystal Name | Geometry/Space Group | Lattice Parameters | Angles |
|---|---|---|---|---|
| **_a_** | $CuFeO_2$ | *Trigonal/ R -3 m :R* | a=b=c=5.96 Å | α=β=γ=29.43° |
| **_b_** | $CuFeO_2$ | *Trigonal/ R -3:H* | a=b=3.0345 Å, c=17.166 Å | α=β=90°, γ=120° |
| **_c_** | $CuFe_5O_8$ | *Cubic/F m -3 m* | a=b=c=8.413 Å | α=β=γ=90° |
| **_d_** | $CuFeO_2$ | *Hexagonal/ P 63 m m c* | a=b=3.035 Å, c=11.449 Å | α=β=90°, γ=120° |
| **_e_** | $CuFe_2O_4$ | *Tetragonal/ I 41/ a m d:2* | a=b=5.8 Å, c=8.73 Å | α=β=γ=90° |
| **_f_** | $CuFe_2O_4$ | *Cubic/F d 3 m* | *a=b=c=8.37 Å* | α=β=γ=90° |

Compared to $LiCrO_2$, the pre-fitting analysis for $CuFeO_2$ was more challenging due to the greater similarity in peak shapes and positions for all calculated spectra. Four of the six candidate crystal structures were subjected to a more detailed pre-fitting analysis by comparing the sum of their calculated EXAFS data with the experimental FT-EXAFS data. This analysis revealed that two CIF files, containing crystal structure information for tetragonal *'I 41/a m d :2'* $CuFe_2O_4$ (Cuprospinel) and cubic *'F d -3 m'* $CuFe_5O_8$, exhibited a high degree of similarity to the experimental data (**Figure 14**). While the researchers had initially anticipated a $CuFeO_2$ structure, the presence of $CuFe_5O_8$ was unexpected.

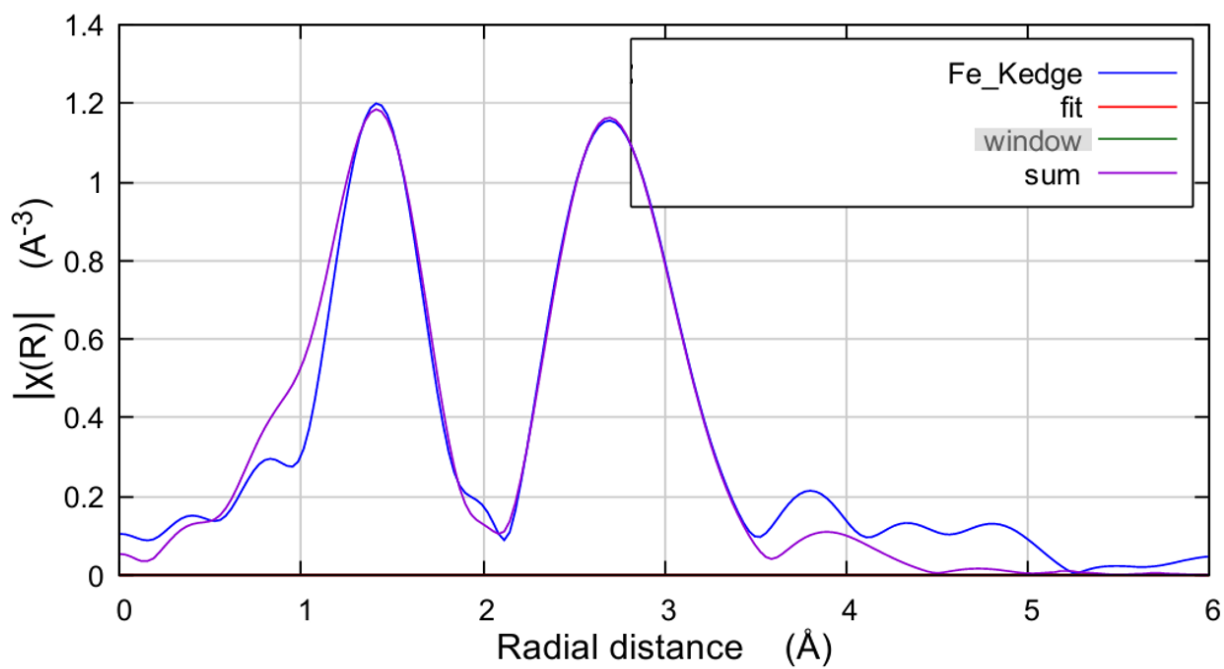

**Figure 14.** Comparison of Experimental FT-EXAFS Data with Calculated (FEFF) Sum of $CuFe_2O_4$ and $CuFe_5O_8$

The comparison of experimental FT-EXAFS data with the calculated sum of $CuFe_2O_4$ and $CuFe_5O_8$ clearly indicates a strong match. Therefore, the next step is to proceed with the fitting process using these two crystal structures. Fitting with these CIF files can be complex, especially when using inaccurate atomic positions. This is because the system initially treats the material as a single crystal. The fitting procedure should prioritize the crystal structure that exhibited the highest similarity during the pre-analysis. $CuFe_2O_4$ data was selected as the starting point for the fitting process. Careful consideration must be given to refining the guess parameters with each iteration, ensuring that previous guesses are discarded and a clear guess frame is established. If the log file indicates high delta R values, suggesting interference from neighboring metal or oxygen atoms, the fitting process should be paused and the crystal card parameters adjusted, maintaining the atomic positions as close as possible to the previous values. If the refined atomic positions do not match the crystal structure in the CIF file when adjusting the lattice parameters, it may indicate the presence of a mixed crystal phase. Similar procedures apply to single-component crystal fitting, but the errors may be less significant due to the simpler system. After each iteration of the fitting process, as described in the previous section, the determined crystal structure parameters should be applied to the CIF file and subsequently used in Rietveld analysis. The excellent agreement between the refined XRD pattern and the experimental data confirms the accuracy of the analysis. **Figure 15** summarizes the key steps of the fitting process and the final results applied to the XRD pattern analysis of $CuFe_2O_4$. Here, the key would be selection of the short range order rather than long range order to obtain best estimation.

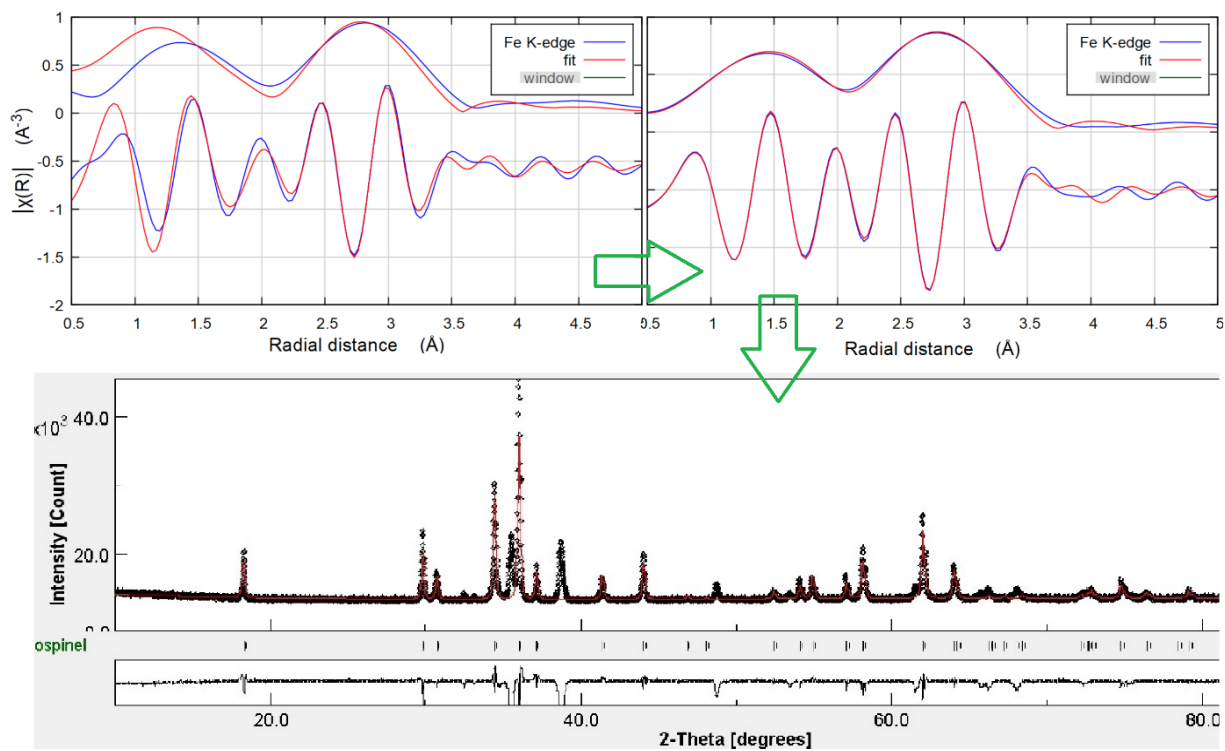

**Figure 15.** Iterative EXAFS Fitting and Rietveld Refinement for $CuFe_2O_4$

The identification of tetragonal $CuFe_2O_4$ as the dominant crystal structure in the sample was unexpected, as researchers had initially believed it to be delafossite $CuFeO_2$. Upon closer inspection, however, some peaks in the XRD pattern were not fully accounted for by the

CuFe$_2$O$_4$ structure. As previously mentioned, a minor phase of CuFe$_5$O$_8$ was identified. A similar analysis was conducted for CuFe$_5$O$_8$, but with fixed parameters for the CuFe$_2$O$_4$ phase to focus on the subtle changes in CuFe$_5$O$_8$. The refined parameters and corresponding Rietveld results are presented in **Figure 16**.

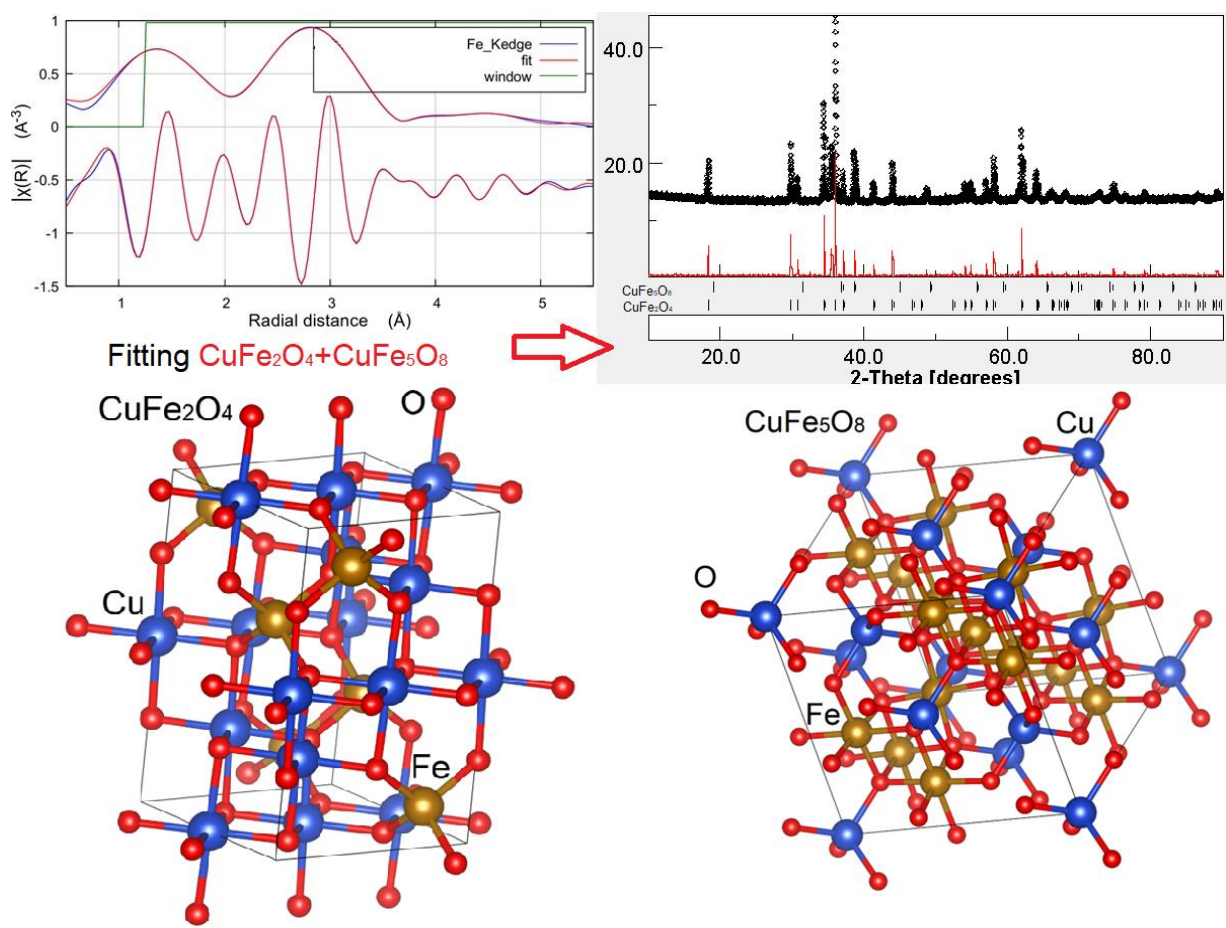

**Figure 16.** Iterative EXAFS Fitting and Rietveld Refinement for CuFe$_2$O$_4$ and CuFe$_5$O$_8$, and Crystal Structures of CuFe$_2$O$_4$ and CuFe$_5$O$_8$

The observed high symmetry and accurate matching of indexed peaks in the XRD pattern validate the accuracy of the IEA technique when coupled with careful parameter selection and meticulous analysis procedures. Inverse EXAFS Analysis Algorithm, as a summary of the given details above is given in **Figure 17**.

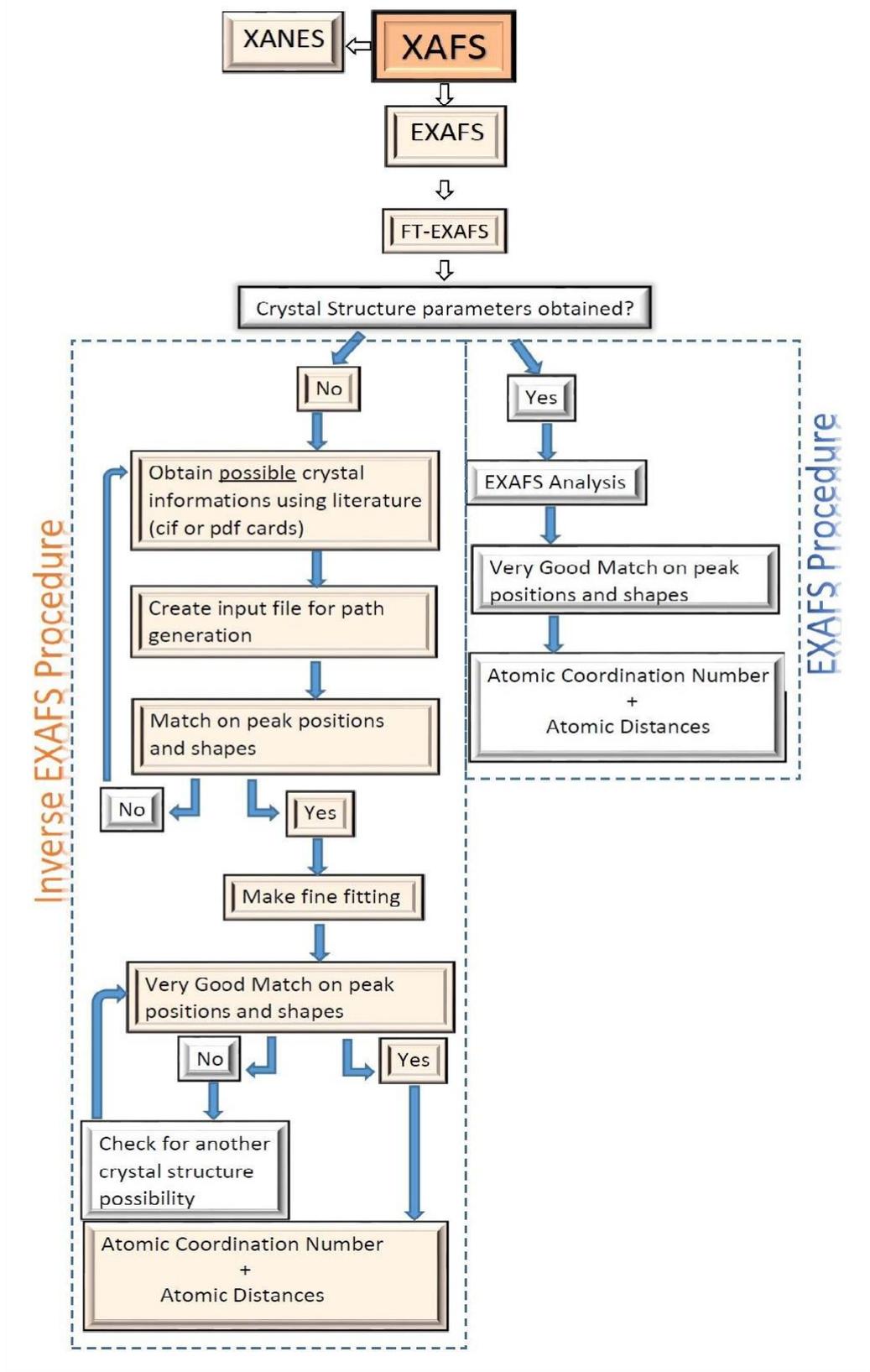

**Figure 17.** Inverse EXAFS Analysis Algorithm

**Conclusions**

A study presents the development of an alternative method for crystal structure analysis, a fundamental and crucial aspect of technological infrastructure. To evaluate the reliability of this new method, Inverse EXAFS Analysis (IEA) was employed to analyze the crystal structures of various experimentally studied materials, preceding the conventional XRD technique. The materials included $LiCrO_2$, where a single crystal dominates the overall structure, and $CuFeO_2$, where the researchers hypothesized the coexistence of two distinct crystal structures as revealed by IEA.

The study demonstrates that Inverse EXAFS Analysis (IEA), a relatively new technique for crystal structure analysis, has the potential to more effectively harness XAFS data for researchers. It offers a reliable alternative to other techniques that can be compromised by instrumentation limitations or crystal structure defects. This research suggests that with further development, IEA can enhance the utility of XAFS, providing researchers with a more robust and user-friendly tool.